\documentclass[aps,pra,twocolumn,groupedaddress,showpacs]{revtex4}
\usepackage{graphicx}
\graphicspath{{figure file fold path}}
\usepackage[dvipsnames,usenames]{color}
\usepackage{color}
\usepackage{float}
\usepackage{subfigure}
\usepackage{amsmath}
\usepackage{amssymb}

\emergencystretch=\maxdimen
\begin{document}
\title{Extreme events generated in microcavity lasers and their predictions by reservoir computing}
\author{T. Wang$^{1,\footnote{twang6@xidian.edu.cn}}$, H. X. Zhou$^{1}$, Q. Fang$^{2}$, Y. N. Han$^{1}$, X. X. Guo$^{1}$, Y. H. Zhang$^{1}$, C. Qian$^{3}$, H. S. Chen$^{3}$, S. Barland$^{4}$, S. Y. Xiang$^{1, \footnote{syxiang@xidian.edu.cn}}$, G. L.~Lippi$^{4,\footnote{Gian-Luca.LIPPI@univ-cotedazur.fr}}$}

\affiliation{$^1$ State Key Laboratory of Integrated Service Networks, Xidian University, Xi’an 710071, China}
\affiliation{$^2$ School of Electronics and Information, Hangzhou Dianzi University, Hangzhou 310018, China}
\affiliation{$^3$ Interdisciplinary Center for Quantum Information, College of Information Science and Electronic Engineering,Zhejiang University, Hangzhou 310027, China}
\affiliation{$^4$ Universit\'e de la C\^ote d'Azur, Institute de Physique de Nice, UMR 7010 CNRS, France}

\begin{abstract}
Extreme events generated by complex systems have been intensively studied in many fields due to their great impact on scientific research and our daily lives. However, their prediction is still a challenge in spite of the tremendous progress that model-free machine learning has brought to the field. 
We experimentally generate, and theoretically model, extreme events in a current-modulated, single-mode microcavity laser operating on orthogonal polarizations, where their strongly differing thresholds -- due to cavity birefringence -- give rise to giant light pulses initiated by spontaneous emission.  Applying reservoir-computing techniques, we identify in advance the emergence of an extreme event from a time series, in spite of coarse sampling and limited sample length.  Performance is optimized through new hybrid configurations that we introduce in this paper.  Advance warning times can reach 5ns, i.e. approximately ten times the rise time of the individual extreme event.

\end{abstract}

\pacs{}
\maketitle

\section{Introduction}

In recent years, Extreme Events (EEs: very rare and large amplitude fluctuations) have been a topic of great interest in various fields due to their significant impact on society~\cite{Cavalcante2013}. While oceanic rogue waves, recognized by their height and steepness well in excess of their average~\cite{Dysthe2008,Pelinovsky2008, Osborne2010}, are the best-known examples, EEs are known to exist, beyond oceanography~\cite{Onorato2001}, in astrophysics~\cite{Hudson1991}, in the atmosphere~\cite{Stenﬂo2009}, in geology~\cite{Bogachev2008}, and even in optics~\cite{Metayer2014, Baronio2012}. Regardless of their different physical origins, these EEs can be seen as emergent phenomena, where information coming from one physical system can benefit the others.  

Optics offers many advantages in this respect, allowing well-controlled laboratory conditions, good detection techniques, as well as extremely fast timescales that allow the collection of large data samples in a limited time.  Powerful instrumentation detects and stores long datasets which may easily contain an isolated EE in the midst of a large number of lower-level oscillations, as shown from experiments and models in nonlinear fibers~\cite{Solli2007,Dudley2008}), lasers subject to external injection~\cite{Bonatto2011,Pisarchik2011}, to self-injection through feedback~\cite{Uy2017,Tlidi2017,Lee2016}, to the presence of an intracavity saturable absorber~\cite{Herink2017,Coulibaly2017,Rimoldi2017}, to soliton interactions~\cite{Walczak2017}, or predicted in resonant parameteric oscillators~\cite{Oppo2013}. 

Paralleling the interest in fundamental understanding of EEs generation in different systems, their prediction is of paramount practical importance, independenly of the physical details.  While model-free machine learning techniques have made significant progress, practical restrictions often limit the amount of information available, increasing the relevance of model-free detection, based on Reservoir Computing (RC), capable of operating with incomplete data with limited accuracy~\cite{Pammi2023}:  a very realistic real-life scenario.

In this paper, we use EEs which appear in the emission of one polarization channel of a low-frequency-pump-modulated semiconductor microcavity laser.  Their observation is corroborated by numerical modelling, based on a standard spin-flip model, which confirm the presence of EEs whose number can be tuned by a suitable choice of modulation parameters.  We show that RC is capable of predicting the emergence of EEs with good accuracy.  Given the broad variety of RC implementations, in order to identify its most successful forms for this task we analyse different schemes and propose several hybrid configurations.  Our work provides a new platform for studying EEs in a microcavity system and beyond, thus opening up a new field of investigation for EEs prediction.

\section{Experimental design}

\begin{figure}
\centering
\includegraphics[width=0.85\linewidth,clip=true]{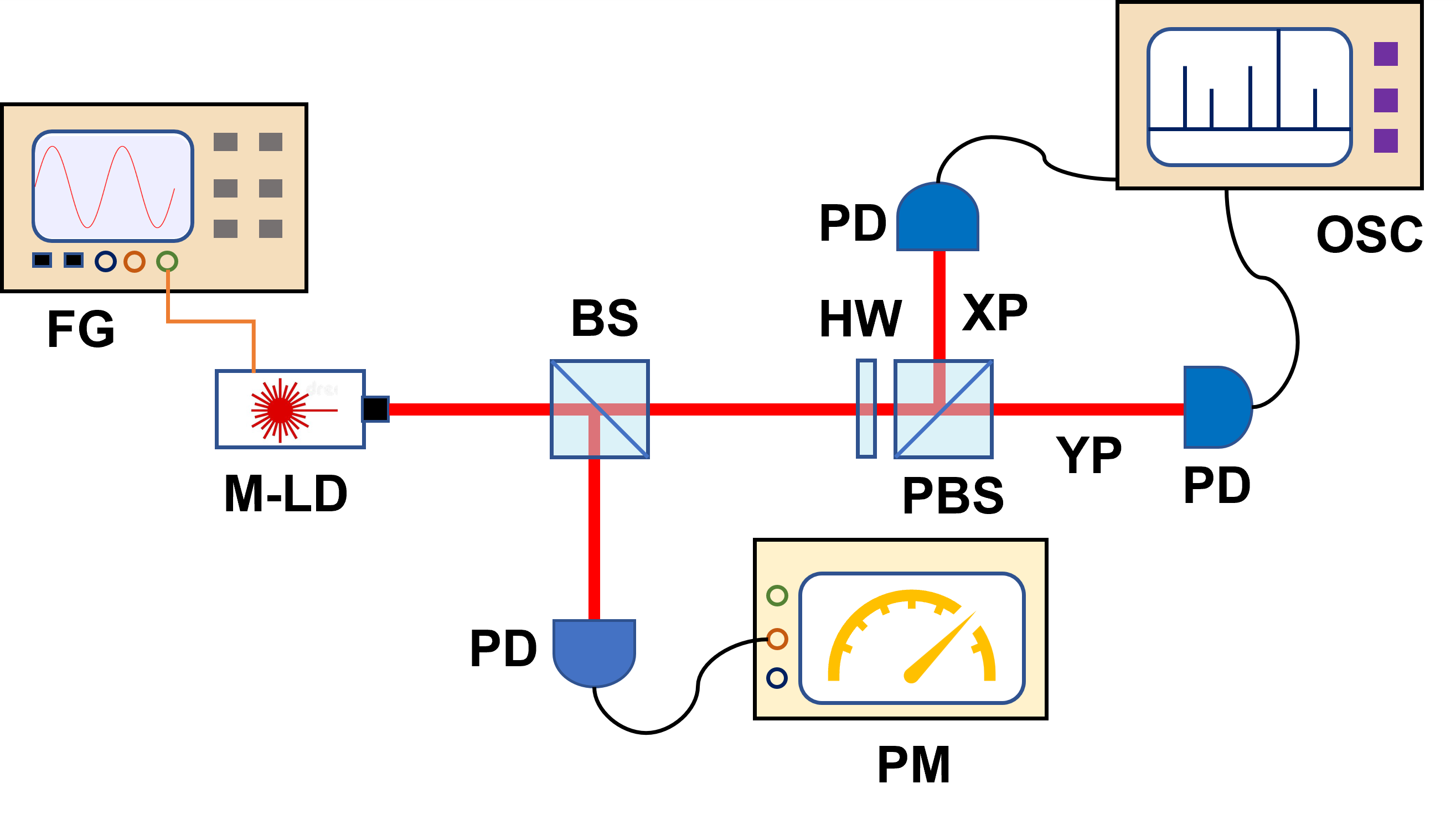}
\caption{
Experimental design: M-LD, microcavity laser diode; OI, Faraday optical isolator; BS, Beam splitter; PD, photodiode; PM, power meter; HW, half-wave plate; PBS, polarization beam splitter; XP, X-polarization; YP, Y-polarization; OSC, oscilloscope; FG, Function generator.
}
\label{SYSTEM}
\end{figure}

The experimental setup is shown in Fig.~\ref{SYSTEM}. The physical system is a single-mode semiconductor microcavity laser diode (M-LD, Ulm Photonics, single mode, $\lambda = (980 \pm 3) {\rm nm}$) where the $\beta$ factor, which represents the fraction of spontaneous emission coupled into the lasing mode, is estimated at $\approx 10^{-3}$~\cite{Wang2015}.  The laser pump is sinusoidally modulated at $\approx 600 {\rm MHz}$ by a function generator (FG, Agilent E4421B), coupled to the stabilized current supply with a bias-tee.  In the following, we will choose a below-threshold bias~\cite{Wang2016,Lippi2022} for the weak field polarization. The output laser beam is then split in a 10:90 ratio after passing though a beam splitter (BS), where the weak beam is detected by a power meter while the other one is analyzed by two channels separated by a Polarized Beam Splitter (PBS), preceded by a Half-Way-Plate (HW) to align the laser's polarization axes with the PBS. The two orthogonal polarization components of the beam are then detected by two fiber-coupled fast photodetectors (Thorlabs PDA8GS). Finally, the temporal signals of the two polarization states are recorded by a digital oscilloscope (LeCroy Wave Master 8600A, sampling rate 10Gs/s, analog bandwidth 6 GHz).

\section{Characterization of temporal EEs}

\begin{figure}
\centering
\includegraphics[width=0.65\linewidth,clip=true]{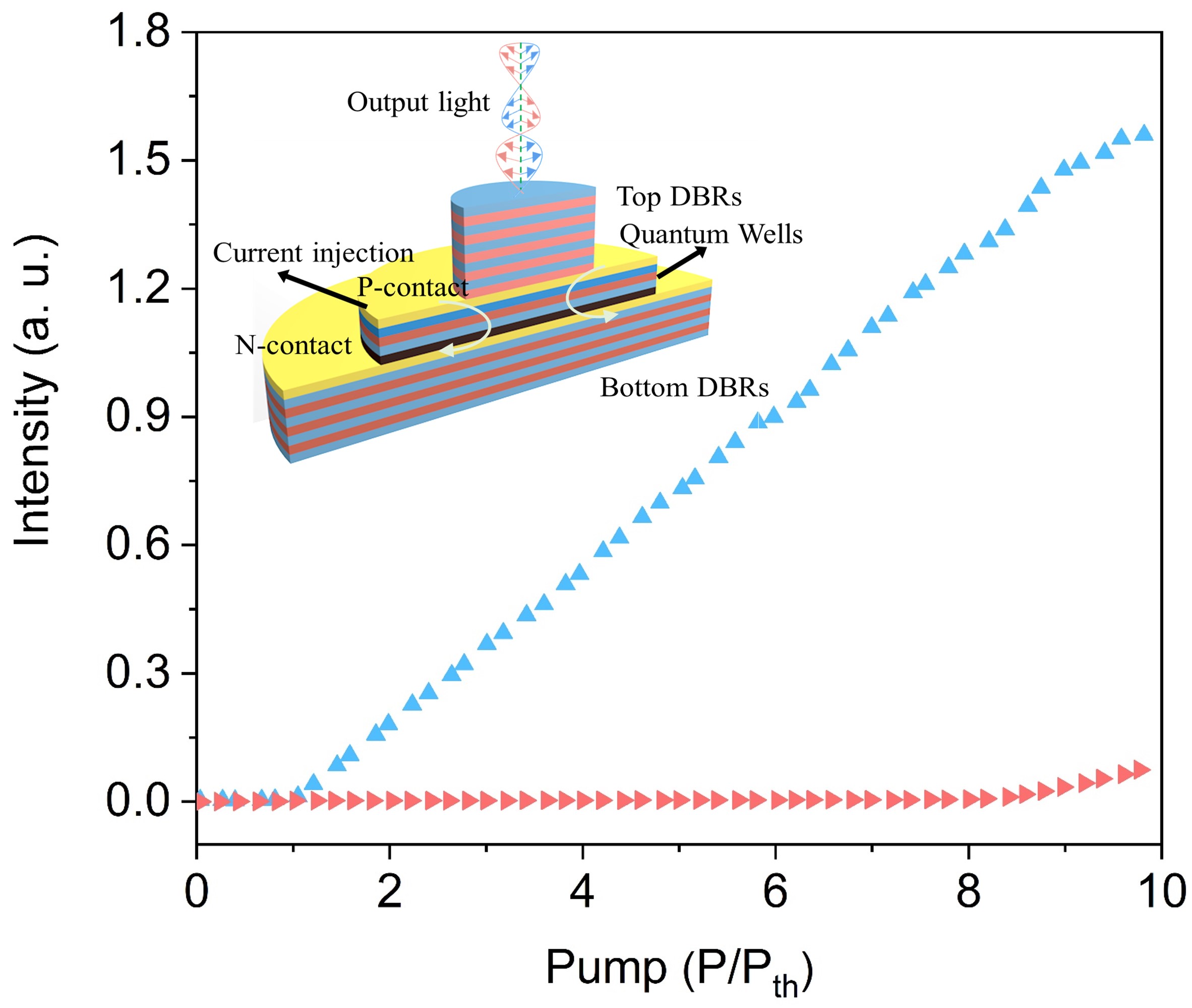}
\caption{
Input-output lasing behaviour curves for TE (blue) and TM (orange) modes of the free running laser: the measured threshold of TE mode is 0.20 mA, and TM mode is around 1.60 mA. Inset: schematic cross-sectional view of vertical cavity laser structure. 
}
\label{Threshold}
\end{figure}

Fig.~\ref{Threshold} shows the measured input-output function curves of the TE (blue) and TM (orange) modes under free-running conditions; the relative pump axis (Pump $(P/P_{th})$) is obtained by dividing the current supplied to the laser by the threshold current of the TE mode ($P^{TE}_{th} \approx$ 0.20 mA). The laser behavior curves show that there is a significant difference in the threshold between TM and TE modes. Specifically, the threshold for the TM mode ($P^{TM}_{th} \approx$ 1.60 mA) is approximately 8 times larger than the one for the TE mode.  We remark that in the broad interval of single-polarization emission, the latter is stable, devoid of the switching behaviour often observed~\cite{Sondermann2004} in Vertical Cavity Surface Emitting Lasers (VCSELs, the schematic representation is shown in the inset of Fig.~\ref{Threshold}).  Thus, in our device the TE polarization entirely controls the microcavity emission, with output which increases linearly with pump. In other words, no deterministic interpolarization dynamics is present in the measured signal.

\begin{figure}
\centering
\includegraphics[width=0.65\linewidth,clip=true]{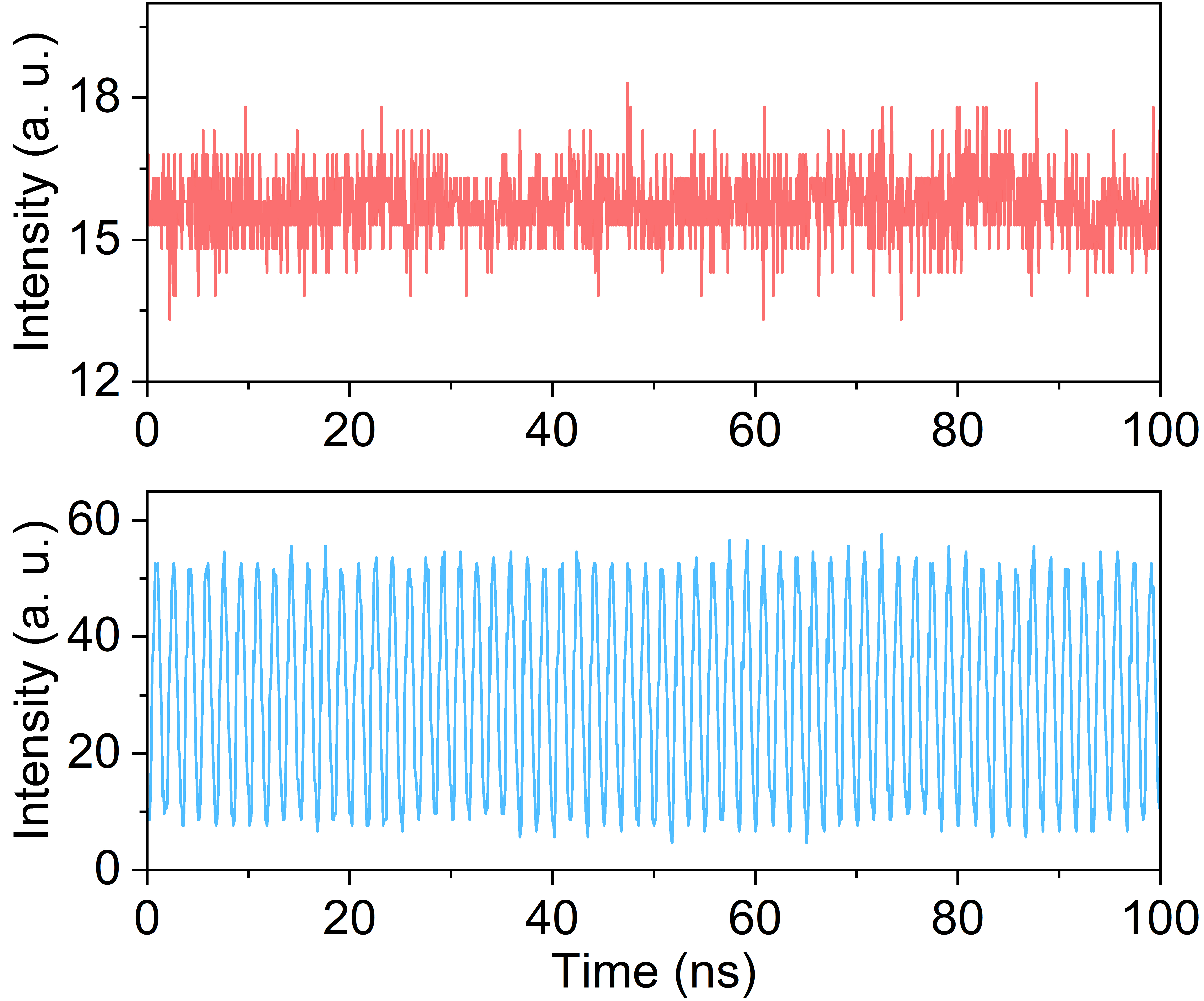}
\caption{
Temporal dynamics  within 100 ns of TE (blue) and TM (orange) modes of the modulated laser. 
}
\label{Dynamics}
\end{figure}

When the laser is modulated at $I_{dc} = 5.5P^{TE}_{th}$ by a 600-MHz sinusoidal signal with a moderate modulation amplitude, such that the TM threshold is not attained, the temporal response of the TE mode exhibits periodic behavior, as shown in Fig.~\ref{Dynamics} (bottom panel). In contrast, the TM mode dynamics remains dominated by spontaneous emission (Fig.~\ref{Dynamics}, top panel), hardly affected by the small pump modulation below threshold (noise from the detection chain mixes in, masking possible residual oscillations).

\begin{figure}
\centering
\includegraphics[width=0.65\linewidth,clip=true]{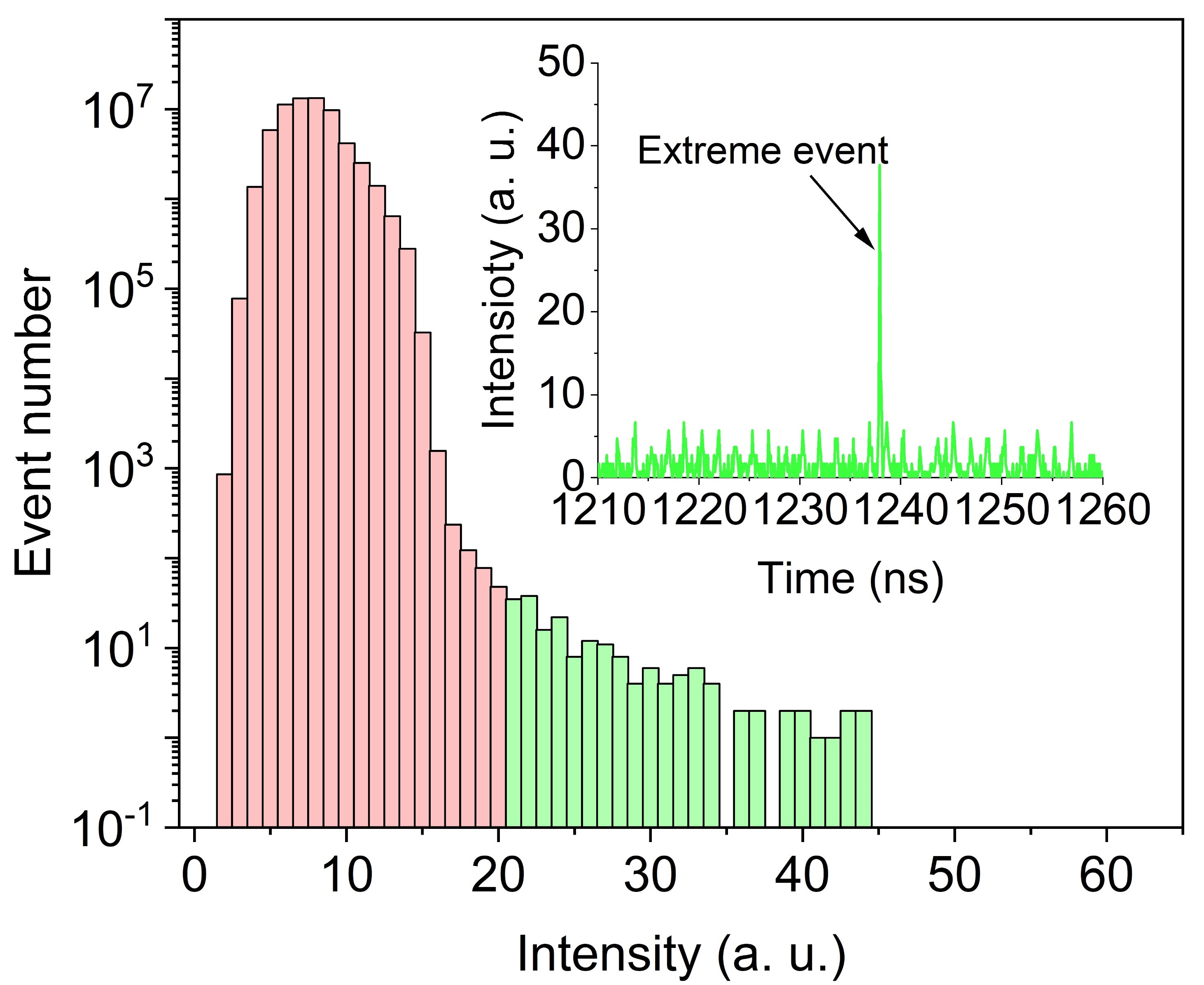}
\caption{
Experimental histograms with EEs, the tail marked by green color represents the events whose intensity is larger than the threshold; inset: typical EE with very large amplitude. 
}
\label{Y-histo}
\end{figure}

Increasing the modulation amplitude to levels which bring the laser, in transient, close to or beyond the TM mode threshold, one observes large amplitude fluctuations in the TM channel.  Fig.~\ref{Y-histo} shows the histogram of the TM mode emission under these conditions, where the long tail at large amplitudes reveals the presence of non-gaussian statistics and the green part of the distribution represents the EEs.  The latter are determined using the standard EE discrimination criterion based on $I_{EE} > \langle I \rangle + 8 \sigma_{I}$~\cite{Bonatto2011}, where $\langle I \rangle$ stands for the average measured intensity and $\sigma_I$ for its standard deviation. The inset of Fig.~\ref{Y-histo} shows a typical rare event in the time domain:  the very large amplitude pulse is surrounded by a background and by smaller spikes which correspond to a regular transient excitation of the TM mode.

\section{Numerical confirmation and analysis}

The experimental observations can be interpreted with the help of the Spin Flip Model (SFM)~\cite{Regalado1997, Valle2008} for a current-modulated semiconductor laser:

\begin{gather}
\label{fields}
\frac{dE_{\pm}}{dt} =  \kappa(1+i\alpha)[(N \pm n)-1]E_{\pm}-(\gamma_a + i\gamma_p)E_{\mp} \notag \\ +\sqrt{\beta (N\pm n)}\xi\pm\, , \\
\label{Carrier1}
\frac{dn}{dt}  =  - \gamma_s n - \gamma[(N+n)\vert E_+ \vert^2 - (N-n)\vert E_- \vert^2]\, , \\
\label{Carrier2}
\frac{dN}{dt}  =  -\gamma [N - I + (N + n)\vert E_+ \vert^2 + (N-n)\vert E_- \vert^2]\, ,
\end{gather}

\noindent where $E_\pm$ are the left and right circularly polarized components of the slowly varying optical field.  $N$ describes the evolution of the total carrier number, in excess of its value at transparency, normalized to the same quantity evaluated at the lasing threshold.  The variable $n$ represents instead the difference between the population carrier numbers which interact with the two different polarizations, normalized in the same way as~\cite{Regalado1997}.  
The decay rate of the field in the cavity is denoted by $\kappa$; $\alpha$ stands for the electric field's phase-amplitude coupling. The carrier relaxation rate is $\gamma$; $\gamma_s$ is the coupling rate between the two circularly polarized radiation channels, coming from the different microscopic relaxation mechanisms that equilibrate the carriers' spin. $\gamma_p$ ($\gamma_a$) represents the degree of linear birefringence (dichroism) per intracavity round-trip time. Essentially, these parameters reflect the degree to which the intracavity radiation is split into two different polarization states. The spontaneous emission factor $\beta$ matches the one already introduced. 
The noise terms $\xi_\pm$ are zero--mean Gaussian white noise terms used to model random fluctuations in the system. 

The normalized injection current $I$ is modulated by a sinusoidal function, which reproduces the experiment, with a constant component $I_{dc}$, a modulation amplitude $\Delta I$ and a modulation frequency $f_m$:  $I(t) = I_{dc} + \Delta I \cos(2 \pi f_m t)$.  The two orthogonal polarization components (TE and TM) have strongly separate thresholds due to cavity birefringence, modelled by the parameter $\gamma_p$.  Their detection is reconstructed through the projection onto the $X$ and $Y$ axes, respectively, computed as:

\begin{gather}
\label{IX}
I_x =  \vert E_x \vert^2 = \frac{1}{2}(E_++E_-)^2\, , \\
\label{Iy}
I_y = \vert E_y \vert^2 = \frac{1}{2}(E_+-E_-)^2\, , 
\end{gather}
where $I_j$ is proportional to the measured intensity in each polarization channel.

\begin{figure}
\centering
\includegraphics[width=0.65\linewidth,clip=true]{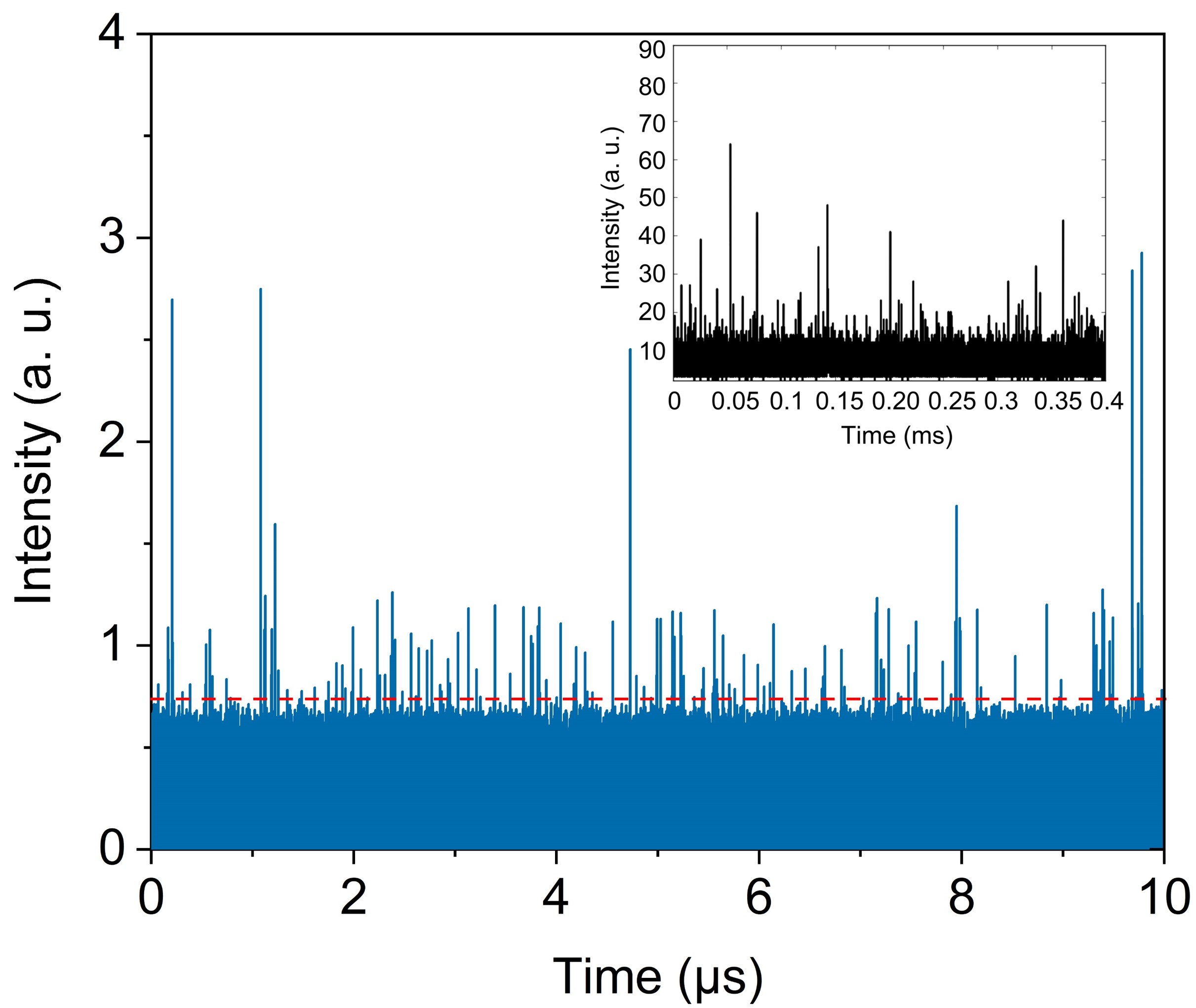}
\caption{
Simulated temporal dynamics of the TM mode in a time interval of 10 $\mu$s:  the red, dashed line indicates the threshold of EEs.  Inset: experimental temporal dynamics of the TM mode in a $0.4 ms$ time interval.
}
\label{Temporal}
\end{figure}

To investigate the parameter region where the extreme events can be found, we perform intensive numerical simulations using the SFM model. 
%
%
After determining the optimal bias position ($I_{dc} = 6.03P^{TE}_{th}$), we proceed to modulate the laser at this specific point~\cite{footnote}.  
Fig.~\ref{Temporal} shows a $10 \mu s$ long snapshot of the TM intensity evolution for $\Delta I = 5.2 P_{th}$ and $f_m = 600 MHz$. The temporal dynamics show the isolated intensity pulses which overcome the threshold for EEs (red line, placed at $I \approx 0.7316$).  For comparison, the inset diplays an experimental sequence.  The satisfactory agreement justifies the use of numerical simulations as a complement in testing RC recognition~\cite{expnum}.
%
%
We remark that EEs strongly depend on $\Delta I$ and $f_m$ in the simulations and appear only in a narrow parameter interval.

The simplicity of the model, which clearly separates into a deterministic and a stochastic part, lends credibility to the action of the spontaneous emission as source of the EEs~\cite{Bonatto2011,Zamora-Munt2013}, since the complexity of an attractor collision is not present in our system in the conditions of the experiment~\cite{Metayer2014}.

\section{Prediction of EEs by RC}\label{sRC}

RC is a novel type of recurrent neural network (RNN)~\cite{Tanaka2019} that has demonstrated exceptional performance in nonlinear channel equalization~\cite{Masaad2022}, time series prediction, and speech recognition~\cite{Hasegawa2023, Jaeger2004, Tanaka2019}. Its unique architecture allows RC to efficiently process complex and high-dimensional data, making it a promising tool for various prediction and classification tasks. Here, our goal is to obtain an accurate statistical prediction for the temporal EEs.  

\begin{figure}
\centering
\includegraphics[width=1\linewidth,clip=true]{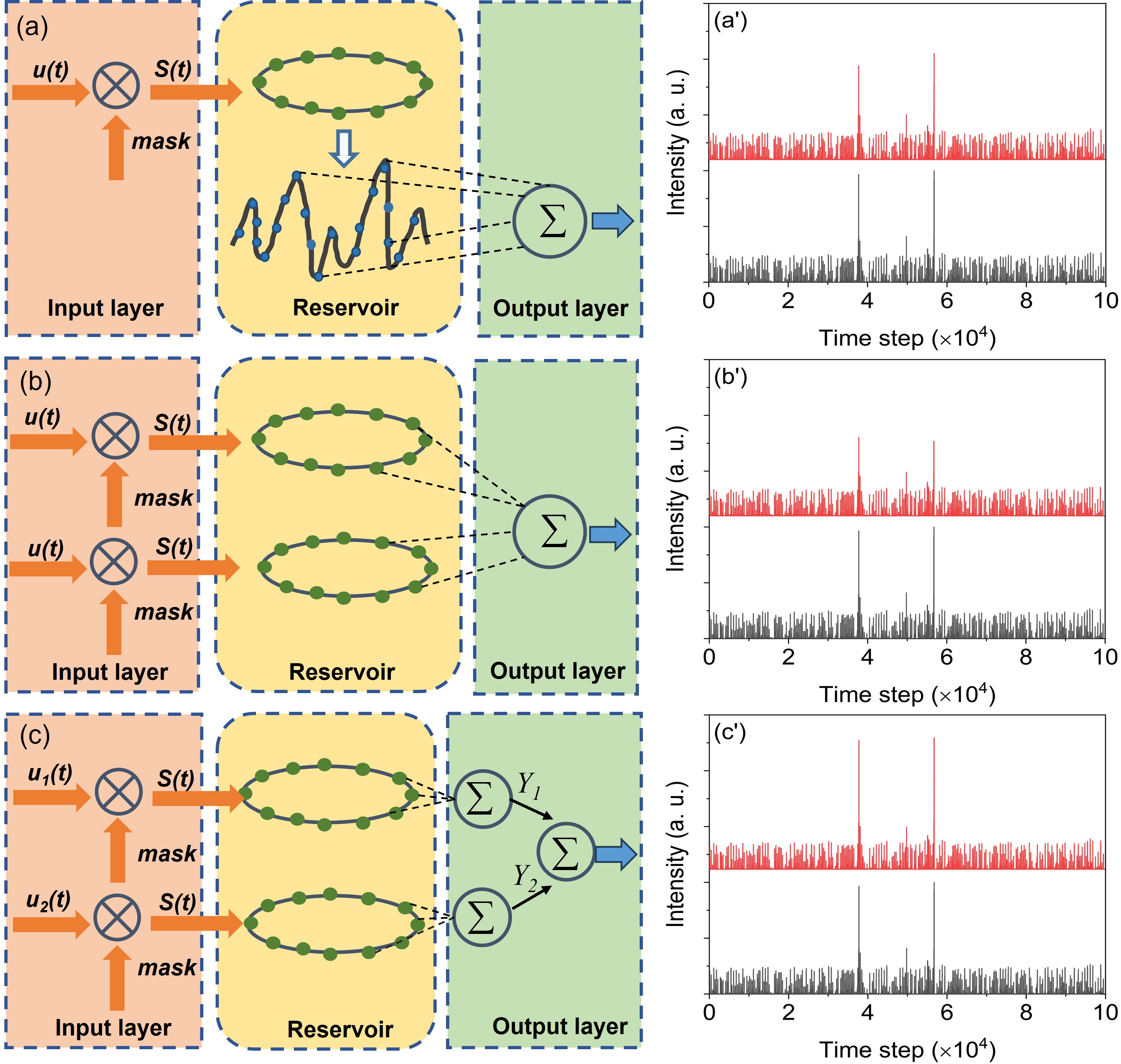}
\caption{
Different prediction configurations: (a) single RC; (b) parallel RC; (c) dual-training RC. 
}
\label{RC-configurations1}
\end{figure}

We numerically implement three different RC configurations (Fig.~\ref{RC-configurations1}) as basic architectures for EE prediction.
The first is a single RC (panel a). Its implementation could also be based on a semiconductor laser (SL) with a feedback loop (FL)~\cite{Brunner2013, Wang2023} as a replacement for the nonlinear nodes (yellow box). The input layer takes an input function signal $u(t)$, which is then multiplied by a mask consisting of randomly assigned integers (-1 and +1) to obtain the signal $S(t)$. The information from the FL is sampled at $\theta$ intervals to collect the high-dimensional information state:  $state(n)$. As a result of the sampling in the FL, the $N$ virtual nodes are generated. The output layer remains the same as in the Echo State Network (ESN).

The second approach, called parallel reservoir configuration, could be implemented with multiple SLs with FL set up in parallel (Fig.~\ref{RC-configurations1}b). Here, $u(t)$ is multiplied by different mask signals $mask_1$ and $mask_2$ to obtain $S_1(t)$ and $S_2(t)$, respectively, which are then injected into different SLs. In the output layer, the virtual node states are used to generate the output signal by calculating a weighted linear sum of the virtual node states. Thus, the total number of virtual nodes is given as $N_{total} = kN$, where $k$ is the number of SLs and $N$ is the number of virtual nodes for each SL. In this scheme, we set $k=2$ for simplicity. The advantages of a parallel configuration are an increase in flow rate and redundancy, which ensure uniterrupted operation also in case of failure of one (or more) nodes. Therefore, this configuration is preferred for applications where high flow rates and increased reliability are essential.

As a third method, we propose a dual-training RC configuration~\cite{Gu2022} (Fig.~\ref{RC-configurations1}c). It consists of an input layer, a storage layer, and an output layer and requires two training processes. In the first, the input $u_1(t)$ is processed in the same way as the single RC.  For the second, the input signal $u_2(t)$ is constructed from the difference between the target values and the predicted values of the first training process. Both $u_1(t)$ and $u_2(t)$ are converted into different mask signals. Finally, the output of the system is the sum of $Y_1$ and $Y_2$.

The performance evaluation of different RCs is based on a data set of 300,000 data points for training and an additional 100,000 points for testing. The virtual nodes of the single, parallel, and dual training RCs were all set to $N_{total} = 100$, while the period of the mask signal was set to a specific value: $T = 2 ps$. In the current work, the feedback delay time was set to $\tau = 1 ns$, the time between the output signal and the input signal, while the sampling interval was set to $\theta = 10 ps$, the time between successive samples. 


Figure~\ref{RC-configurations1}a'-c' displays the results of the EE prediction, after training, for each RC. The original (target) signal is in black, while the red curve shows the predicted signal. There is a close resemblance in all cases, albeit some differences in amplitudes. Overall, RCs appear to be a promising tool in forecasting EEs.  A closer analysis provides a better insight into RC performance.

\begin{figure}
\centering
\includegraphics[width=1\linewidth,clip=true]{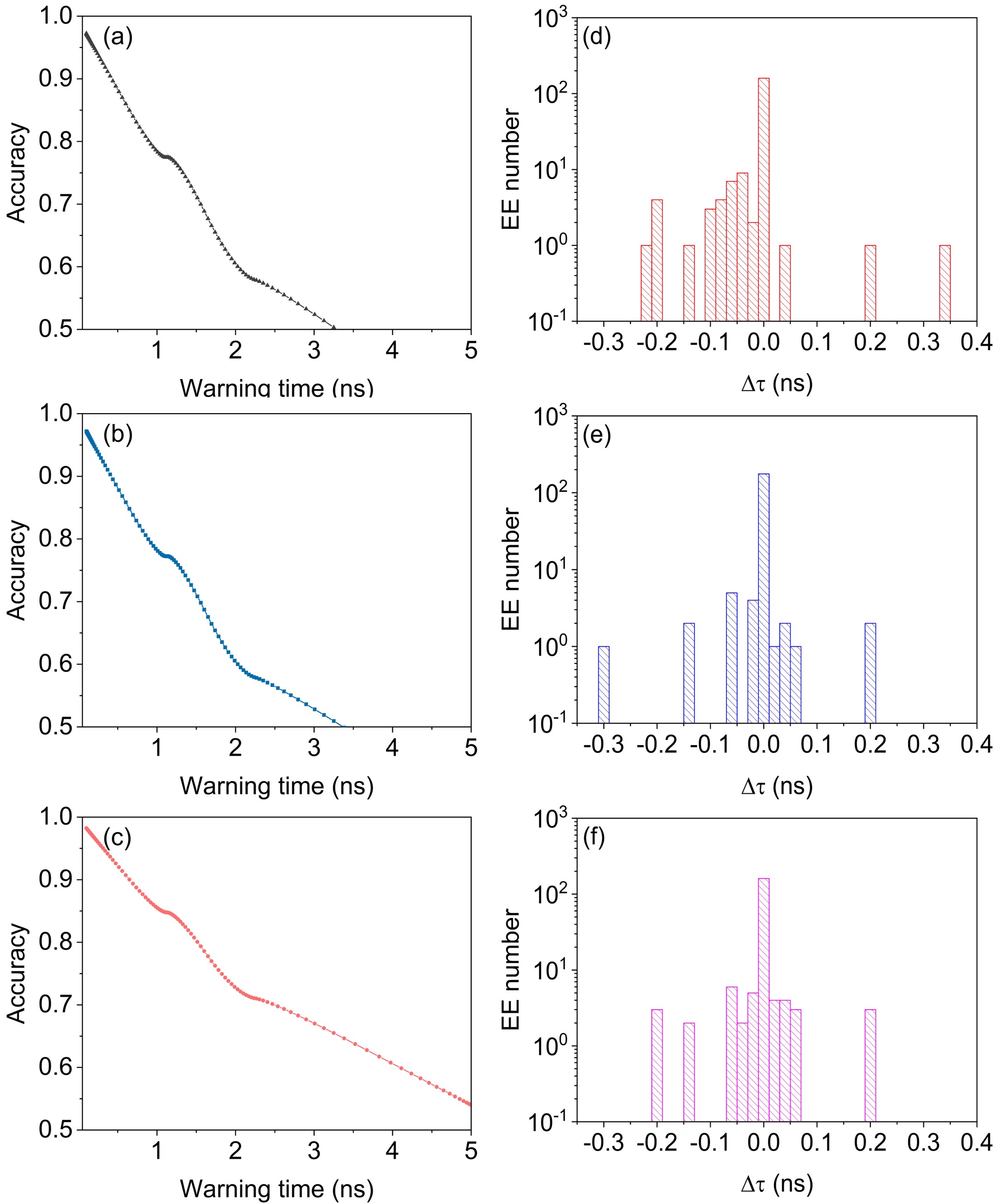}
\caption{Calculated forecasting accuracy (a-c) and histograms of error distributions (d-f) with different RCs configurations: (a) and (d), single RC; (b) and (e), parallel RC; (c) and (f), Dual-training RC.}
\label{Accuracy-time}
\end{figure}

Fig.~\ref{Accuracy-time}a-c shows the calculated prediction accuracy of EEs over time for the three different RCs. Single and parallel RCs have comparable performance. Their prediction accuracy is above 0.9 for shorter warning times, but starts to rapidly decrease after $\approx 1 {\rm ns}$. A short plateau appears, followed by further decline which reaches 50\% accuracy when the warning time reaches $\approx 3.25 {\rm ns}$. The dual-training RC, instead, proves to be much more efficient in terms of overall performance. In fact, its accuracy remains abot 50\% even when the warning time is pushed to the $5 {\rm ns}$.  

Finally, to evaluate the accuracy of the different RC configurations in predicting the time of appearance of EEs, we calculated the errors between the original and predicted time positions and plotted the distributions for each error value. As shown in Fig.~\ref{Accuracy-time}d-e, the horizontal axis represents the deviation values in units of time, measured in ns. The vertical axis shows the number of EEs corresponding to each error value, thus offering a quantitative and visual measure of the timing errors in predicting an EE.  In addition, a higher concentration of data points at or near zero time deviation signals a higher accuracy level, while a scattered distribution of data points indicates a poorer one.

For the single RC, even though the highest column is at position $\Delta \tau = 0$, there are still several errors, with $\Delta \tau$ reaching values as large as $0.338 {\rm ns}$ (Fig.~\ref{Accuracy-time}d). This suggests that while single RC may be accurate for some aspects of the task, there is still significant room for improvement. For the parallel RC, $\vert \Delta\tau \vert$ further regroups around  $0$, indicating a better temporal prediction with a maximum deviation of $0.3 {\rm ns}$ (Fig.~\ref{Accuracy-time}e); a marked improvement over the single RC.  Fig.~\ref{Accuracy-time}f shows that the majority of errors in the dual-training RC occur in the range of $\vert \Delta\tau \vert < 0.1 ns$, indicating a higher level of accuracy than the other two RCs, and a maximum deviation $0.2 ns$. Thus, the dual-training RC offers superior prediction capabilities and is more effective for accurate EEs prediction.

\section{Discussion and conclusions}

The analysis has yielded EE predictions that, in our experimental and numerical data, exceed 50\% accuracy for the dual-training RC network at times as long as 5 ns.  To put this in context and to evaluate the performance more generally, it is important to compare it to the intrinsic timescale of the EE, i.e., its rise-time.  Estimating the rise time as the time taken for the signal to reach its maximum, starting from 10\% of its height (to eliminate the influence of the background), we find a typical rise time of $\tau_{rt} \approx 0.5 ns$.  In relative terms, this means that the dual training network is able to predict an EE with $\tau_{f,0.7} \approx$ 70\% accuracy with a warning time of $\tau_f \approx 5 \tau_{rt}$ (Fig.~\ref{Accuracy-time}f).  This time increases to $\tau_{f,0.5} > 10 \tau_{rt}$ when the accepted accuracy is 50\%.  For events whose main characteristic is the intensity of the phenomenon (e.g. water waves for oil rigs or ships) such an advance would amount to a few minutes warning.

\begin{figure}
\centering
\includegraphics[width=0.65\linewidth,clip=true]{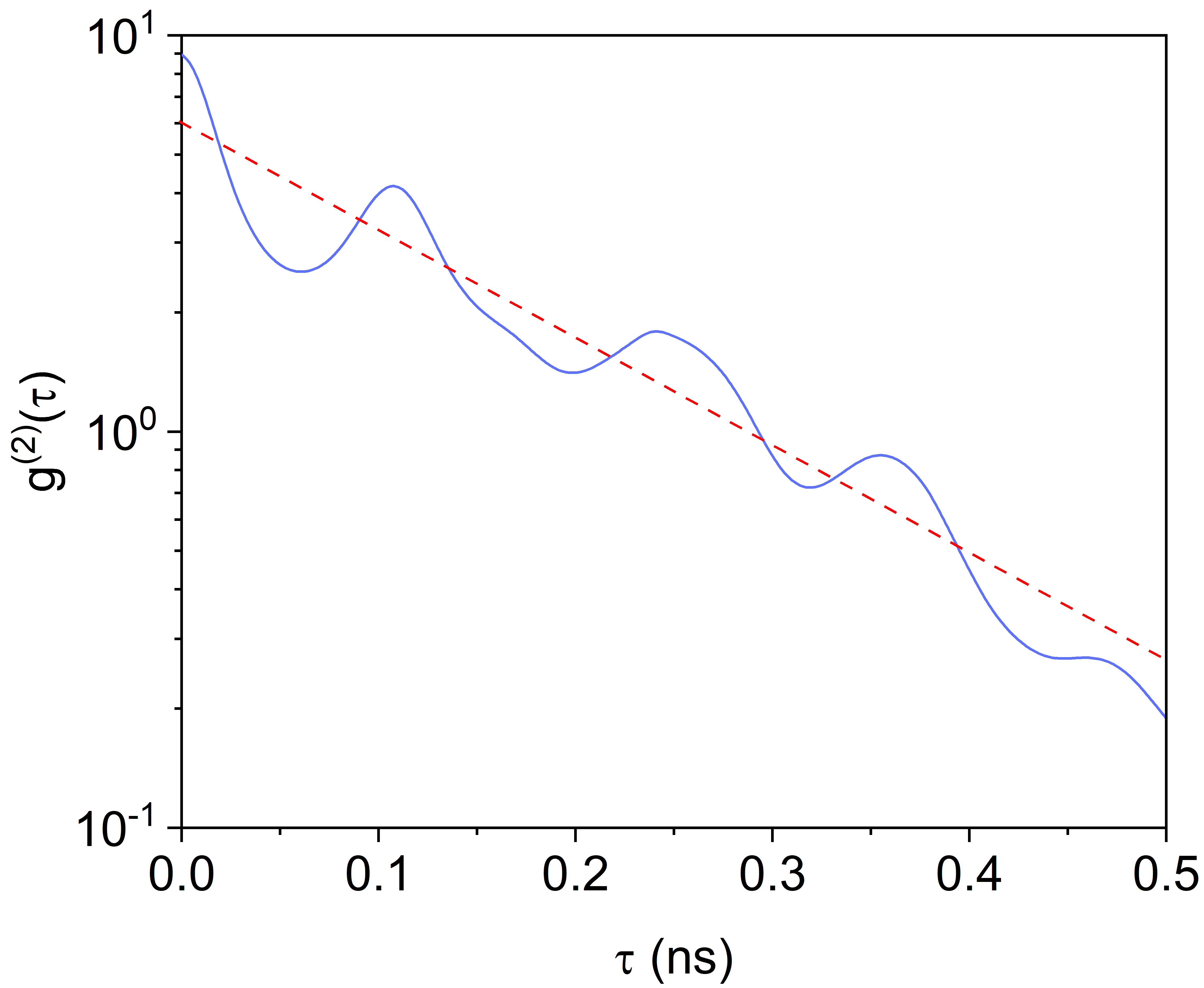}
\caption{Computed second-order, time-delayed autocorrelation on a temporal time sequence containing EEs (solid, blue line). Best fit for the autocorrelation (red, dashed line), with slope $a = -6.321 \times 10^9 s^{-1}$.}
\label{corrtime}
\end{figure}

Another way to quantify the amount of time warning is to compare the advance time to the autocorrelation time of the signal.  Fig.~\ref{corrtime} shows the fit, on a semi-logarithmic scale, of the time-delayed second-order autocorrelation $\left( g^{(2)}(\tau) = \frac{\langle s(t) s(t-\tau) \rangle}{\langle s(t) \rangle^2}. \right)$.  The slope of the line corresponds to a correlation time of $\tau_c \approx 0.16 ns$.  The autocorrelation (solid line, Fig.~\ref{corrtime}) shows a correlation recovery at $\tau \approx 0.1 ns$ and a much smaller one at $\tau \approx 0.25 ns$.  Comparing with the 70\% accuracy of the warning time, we find $\tau_{f,0.7} \approx 14 \times \tau_c$; $\tau_{f,0.7}$ is still about ten times longer than the second revival in the correlation.  Reading from Fig.~\ref{Accuracy-time}f the 90\% accuracy time, we can estimate $\tau_{f,0.9} \approx 5 \times \tau_c$.
These numbers are important as a relative measure of the prediction performance of the dual training RC.  In fact, prediction times up to 5 times the correlation time are a very valuable tool for predicting extreme phenomena~\cite{Jacques-Dumas2022}.   

In conclusion, we have experimentally generated EEs from a vertical microcavity semiconductor laser emitting in a dominant polarization mode (TE) with the pump biased below the threshold of the weak polarization mode (TM), and modulated in time.  The EE arise from the rare strong amplification of a fluctuation (spontaneous emission) that occurs close to the pump maximum, favoring the dominance of the TM polarization.  Using the spin-flip model for semiconductor lasers, we have reproduced high-resolution data for the EE, with features consistent with those of the experiment.  The EE prediction has been obtained with the help of Reservoir Computing (RC), for which we have numerically implemented three schemes.  The dual-training RC proves to be the best in terms of both advance warning time and error distribution, and works well on both experimental and numerical data.  The advance warning exceeds five characteristic times and can reach $\approx$ 15 times, depending on the criteria adopted, as explained above.

The growing importance of EEs and early warning, not only in basic research~\cite{Wolde1997,Chandler2005,Kohn2005}, but also in the prediction of natural phenomena and disasters (extreme heat waves~\cite{Jacques-Dumas2022}, extreme precipitation~\cite{Zhang2023}, tropical cyclones or hurricanes~\cite{Bi2023}), gives great relevance to the investigation of techniques for their optimization.  Optical systems, such as microcavity lasers, offer the advantage of flexibility and ease of measurement, the potential to accumulate large amounts of data in a short time, and laboratory controlled conditions.

\section*{Acknowledgment}
This work is partially supported by several sources: National Key Research and Development Program of China (2021YFB2801900, 2021YFB2801901, 2021YFB2801902, 2021YFB2801904), National Natural Science Foundation of China (Grant No. 61804036), Zhejiang Province Commonweal Project (Grant No. LGJ20A040001).


\begin{thebibliography}{90}
\bibitem{Cavalcante2013} H. L. D. de S. Cavalcante, M. Oria, D. Sornette, E. Ott and D.J. Gauthier, ``Predictability and suppression of extreme events in a chaotic system,'' \textit{Phys. Rev. Lett.}, \textbf{111}, 198701, (2013).

\bibitem{Dysthe2008} K. Dysthe, H. E. Krogstad, and P. Muller, ``Oceanic rogue waves,'' \textit{Annu. Rev. Fluid Mech.} \textbf{40}, 287, (2008).

\bibitem{Pelinovsky2008} E. Pelinovsky and C. Kharif, \textit{Extreme Ocean Waves} (Springer, Berlin, 2008).

\bibitem{Osborne2010} A. R. Osborne, \textit{Nonlinear Ocean Waves and the Inverse Scattering Transform}, (Elsevier, New York, 2010).

\bibitem{Onorato2001} M. Onorato, A. R. Osborne, M. Serio, and S. Bertone, ``Freak waves in random oceanic sea states,'' \textit{Phys. Rev. Lett.} \textbf{86}, 5831, (2001).

\bibitem{Hudson1991} H. S. Hudson, ``Solar flares, microflares, nanoflares, and coronal heating,'' \textit{Solar Phys.} \textbf{133}, 357-369, (1991).

\bibitem{Stenﬂo2009} L. Stenflo and P.K. Shukla, ``Nonlinear acoustic-gravity waves,'' \textit{J. Plasma Phys.}, \textbf{75}, 841-847 (2009).

\bibitem{Bogachev2008} M. I. Bogachev, J. F. Eichner and A. Bunde, ``On the occurrence of extreme events in long-term correlated and multifractal data sets,'' \textit{Pure appl. geophys.}, \textbf{165}, 1195-1207, (2008).

\bibitem{Metayer2014} C. Metayer, F. M. de Aguiar, A. Serres, E.J. Rosero, J. R. Rios Leite, W. A. S. Barbosa and J. R. Tredicce, ``Extreme events in chaotic lasers with modulated parameter,'' \textit{Opt. Express}, \textbf{22}, 17850-17859, (2014).	

\bibitem{Baronio2012} F. Baronio, A. Degasperis, M. Conforti, and S. Wabnitz, ``Solutions of the vector nonlinear schr\"odinger equations: evidence for deterministic rogue waves,'' \textit{Phys. Rev. Lett.}, \textbf{109}, 044102, (2012).

\bibitem{Solli2007} D. R. Solli, C. Ropers, P. Koonath and B. Jalali, ``Optical rogue waves,'' \textit{Nature}, \textbf{450}, 1054-1057, (2007).

\bibitem{Dudley2008} J. M. Dudley, G. Genty, and B. Eggleton, ``Harnessing and control of optical rogue waves in supercontinuum generation,'' \textit{Opt. Express}, \textbf{16}, 3644-3651, (2008).

\bibitem{Bonatto2011} C. Bonatto, M. Feyereisen, S. Barland, M. Giudici, C. Masoller, J. R. Rios Leite and J. R. Tredicce, ``Deterministic optical rogue waves,'' \textit{Phys. Rev. Lett.}, \textbf{107}, 053901, (2011).

\bibitem{Pisarchik2011} A. N. Pisarchik, R. J. Re a tegui, R. S. Escoboza, G. H. Cuellar and M. Taki, ``Rogue waves in a multistable system,'' \textit{Phys. Rev. Lett.}, \textbf{107}, 274101, (2011).

\bibitem{Uy2017} C. H. Uy, D. Rontani, and M. Ssciamanna, ``Vectorial extreme events in VCSEL polarization dynamics,'' \textit{Opt. Lett.}, \textbf{42}, 2177-2180, (2017). 

\bibitem{Tlidi2017} M. Tlidi and K. Panajotov, ``Two-dimensional dissipative rogue waves due to time-delayed feedback in cavity nonlinear optics,'' \textit{Chaos}, \textbf{27}, 013119, (2017).

\bibitem{Lee2016} M. W. Lee, F. Baladi, J. R. Burie, M. A. Bettiati, A. Boudrioua, A. P. A. Fischer, ``Demonstration of optical rogue waves using a laser diode emitting at 980 nm and a fiber Bragg grating,'' \textit{Opt. Lett}, \textbf{41}, 4476-4479, (2016).

\bibitem{Herink2017} G. Herink, F. Kurtz, B. Jalali, D.R. Solli, C. Ropers, ``Real-time spectral interferometry probes the internal dynamics of femtosecond soliton molecules,'' \textit{Science}, \textbf{356}, 50-53, (2017).

\bibitem{Coulibaly2017} S. Coulibaly, M. G. Clerc, F. Selmi, and S. Barbay, ``Extreme events following bifurcation to spatiotemporal chaos in a spatially extended microcavity laser,'' \textit{Phys. Rev. A}, \textbf{95}, 023816, (2017).

\bibitem{Rimoldi2017} C. Rimoldi, S. Barland, F. Prati, and G. Tissoni, ``Spatiotemporal extreme events in a laser with a saturable absorber,'' \textit{Phys. Rev. A}, \textbf{95}, 023841, (2017). 

\bibitem{Walczak2017} P. Walczak, C. Rimoldi, F. Gustave, L. Columbo, M. Brambilla, F. Prati, G. Tissoni, and S. Barland, ``Extreme events induced by collisions in a forced semiconductor laser,'' \textit{Opt. Lett.}, \textbf{42}, 3000-3003, (2017).

\bibitem{Oppo2013} G. L. Oppo, A. M. Yao, and D. Cuozzo, ``Self-organization, pattern formation, cavity solitons, and rogue waves in singly resonant optical parametric oscillators,'' \textit{Phys. Rev. A}, \textbf{88}, 043813, (2013).

\bibitem{Pammi2023} V. A. Pammi, M.G. Clerc, S. Coulibaly, S. Barbay, ``Extreme events prediction from nonlocal partial information in a spatiotemporally chaotic microcavity laser,'' \textit{Phys. Rev. Lett.}, \textbf{130}, 223801, (2023).

\bibitem{Wang2015} T. Wang, G.P. Puccioni, and G.L. Lippi, ``Dynamical Buildup of Lasing in Mesoscale Devices'', \textit{Sci. Rep.} \textbf{5}, 15858 (2015); the Supplementary Material explains the measurement technique.

\bibitem{Wang2016} T. Wang, G.P. Puccioni, and G.L. Lippi, ``How mesoscale lasers can answer fundamental questions related to nanolasers'', PRoc. SPIE 9884, 98840B (2016).

\bibitem{Lippi2022}
G. L. Lippi, T. Wang, G. P.Puccioni, ``Why standard threshold definitions fail for nanolasers,'' \textit{Chaos, Solitons and Fractals}, \textbf{157}, 111850, (2022).

\bibitem{Sondermann2004} M. Sondermann, T. Ackemann, S. Balle, J. Mulet, and K. Panajotov, ``Experimental and theoretical investigations on elliptically polarized dynamical transition states in the polarization switching of vertical-cavity surface-emitting lasers'', \textit{Optics Commun.} \textbf{235}, 421-434 (2004).

\bibitem{Regalado1997}
J. Martin-Regalado, F. Prati, M. San Miguel, and N. B. Abraham, ``Polarization properties of vertical-cavity surface-emitting lasers,'' \textit{IEEE J. Quantum Electron.}, \textbf{33}, 765-783, (1997).

\bibitem{footnote} We verify the quality of the match between model and experimental observations by comparing the predicted and observed dynamics.  The satisfactory comparison justifies the use of the modelling approach.

\bibitem{Zamora-Munt2013} J. Zamora-Munt, B. Garbin, S. Barland, M. Giudici, J.R. Rios Leite, C. Masoller, and J.R. Tredicce, ``Rogue waves in optically injected lasers: Origin, predictability, and suppression'', \textit{Phys. Rev. A} \textbf{87}, 035802 (2013).


\bibitem{Valle2008} 
A. Valle, M. Sciamanna, and K. Panajotov, ``Irregular pulsating polarization dynamics in gain-switched vertical-cavity surface-emitting lasers,'' \textit{IEEE J. Quantum Electron.}, \textbf{44}, 136-143, (2008).

\bibitem{expnum}  The scheme discussed in Section~\ref{sRC} successfully predicts EEs both with experimental and numerical data.  In the presentation we concentrate on the numerically generated sequences, which offer better time resolution than the experimental ones, limited by the sampling resolution of the detection chain.

\bibitem{Tanaka2019}
G. Tanaka, T. Yamane, J. B. H\'eroux, R. Nakane, N. Kanazawa, S. Takeda, H. Numata, D. Nakano, A. Hirose, ``Recent advances in physical reservoir computing: a review,'' Neural Netw.,vol. 115, pp. 100-123, 2019.

\bibitem{Masaad2022} S. Masaad, E. Gooskens, S. Sackesyn, J. Dambre and P. Bienstman, ``Photonic reservoir computing for nonlinear equalization of 64-QAM signals with a Kramers-Kronig receiver,'' \textit{Nanophotonics}, \textbf{12}, 925-935, (2023).

\bibitem{Hasegawa2023}
H. Hasegawa, K. Kanno and A. Uchida, ``Parallel and deep reservoir computing using semiconductor lasers with optical feedback,'' \textit{Nanophotonics}, \textbf{12}, 869-881, (2023).

\bibitem{Jaeger2004}
H. Jaeger and H. Hass, ``Harnessing nonlinearity: predicting chaotic systems and saving energy in wireless communication,'' \textit{Science}, \textbf{304}, 78-80, (2004).

\bibitem{Brunner2013} D. Brunner, M.C. Soriano, C.R. Mirasso, and I. Fischer, ``Parallel photonic information processing at gigabyte per second data rates using transient states'', \textit{Nat. Commun.} \textbf{4}, 1364 (2013).

\bibitem{Wang2023}
T. Wang, C. Jiang, Q. Fang, X. Guo, Y. Zhang, C. Jin, S. Xiang, ``Reservoir computing and task performing through using high-$\beta$ lasers with delayed optical feedback,''  arXiv:2305.11878v2, (2023).

\bibitem{Gu2022} B. L. Gu, S. Y. Xiang, X. X. Guo, D. Z. Zheng, and Y. Hao, ``Enhanced prediction performance of a time-delay reservoir computing system based on a VCSEL by dual-training method,'' \textit{Opt. Express}, \textbf{30}, 30779-30790, (2022).

\bibitem{Wolde1997} P. R. Wolde and D. Frenkel, ``Enhancement of protein crystal nucleation by critical density fluctuations'', \textit{Science} \textbf{277}, 1975-1978 (1997).

\bibitem{Chandler2005} D. Chandler, ``Interfaces and the driving force of hydrophobic assembly'', \textit{Nature} \textbf{437}, 640-647 (2005).

\bibitem{Kohn2005} R. V. Kohn, M.G. Reznikoff, and E. Van den Eijnden, ``Magnetic elements at finite temperature and large deviation theory'', \textit{J. Nonlinear Sci.} \textbf{15}, 223-253 (2005).

\bibitem{Jacques-Dumas2022} V. Jacques-Dumas, F. Ragone, P. Borgnat, P. Abry, and F. Bouchet, ``Deep Learning-Based Extreme Heatwave Forecast'', \textit{Front. Clim.} \textbf{4}, 789641 (2022).

\bibitem{Zhang2023} Y. Zhang, M. Long, K. Chen, L. Xing, R. Jin, M.I. Jordan, and J. Wang, ``Skilful nowcasting of extreme precipitation with NowcastNet'', \textit{Nature}, (2023). https://doi.org/10.1038/s41586-023-06184-4

\bibitem{Bi2023} K. Bi, L. Xie, H. Zhang, X. Chen, X. Gu, and Q. Tian, ``Accurate medium-range global weather forecasting with 3D neural networks'', \textit{Nature}, (2023).  https://doi.org/10.1038/s41586-023-06185-3


\end{thebibliography}
\end{document}